\documentclass{article}

\usepackage{arxiv}

\usepackage[utf8]{inputenc} 
\usepackage[T1]{fontenc}    
\usepackage{hyperref}       
\usepackage{url}            
\usepackage{booktabs}       
\usepackage{amsfonts}       
\usepackage{nicefrac}       
\usepackage{microtype}      
\usepackage{lipsum}		
\usepackage{graphicx}
\usepackage{doi}
\usepackage{graphicx}
\usepackage{dcolumn}
\usepackage{bm}
\usepackage{amsmath}
\usepackage{amssymb}
\usepackage{mathptmx}
\usepackage{relsize}
\usepackage{nicefrac}
\usepackage{etoolbox}
\usepackage[inkscapelatex=false]{svg}
\newcommand{\round}[1]{\ensuremath{\lfloor#1\rceil}}

\title{Detecting local perturbations of networks in a latent hyperbolic embedding space}

\author{
    {Alice Longhena,$\quad$ Martin Guillemaud} \\
	CNRS-UMR7225, Inserm-U1127, Sorbonne University-UM75, Inria-Paris (Nerv-Team) \\
    Paris Brain Institute, H\^opital de la Piti\'e Salp\^etri\`ere. 75013 Paris, France\\
    \texttt{alice.longhena@icm-institute.org} \\
    \And
	{Mario Chavez} \\
    CNRS UMR 7225 \\
    H\^opital de la Piti\'e Salp\^etri\`ere. 75013 Paris, France\\
}

\renewcommand{\headeright}{}
\renewcommand{\undertitle}{}

\hypersetup{
pdftitle={Detecting local perturbations of networks in a latent hyperbolic embedding space},
pdfsubject={q-bio.QM, q-bio.NC, physics.data-an},
pdfauthor={Alice Longhena, Martin Guillemaud, Mario Chavez},
pdfkeywords={Brain Networks, Geometric embedding, Epilepsy},
}

\begin{document}
\maketitle

\begin{abstract} 
Graph theoretical approaches have been proven to be effective in characterizing connected systems, as well as quantifying their dysfunction due to perturbation. In this paper, we show the advantage of a non-Euclidean (hyperbolic) representation of networks to identify local connectivity perturbations and characterize the induced effects on a large scale. We propose two perturbation scores based on representations of the networks in a latent geometric space, obtained through an embedding onto the Poincaré disk. 
We numerically demonstrate that these scores are able to localize perturbations in networks with homogeneous or heterogeneous degree connectivity. We apply this framework to identify the most perturbed brain areas in epileptic patients following surgery. This study is conceived in the effort of developing more powerful tools to represent and analyze brain networks, and it is the first to apply geometric network embedding techniques to the case of epilepsy.
\end{abstract}
 
\maketitle

\section{\label{sec:level1} Introduction}
Any system whose behavior can be seen as resulting from the interaction of a set of interconnected units can be modeled in the mathematical form of a graph. The last two decades have seen a growing interest in applying graph theory to describe the complex structure and function of real-world interconnected systems, such as the Internet, social networks, and the brain. Evidence has revealed~\cite{boccaletti2006complex} that these systems share some peculiar characteristics, like the small-world property~\cite{Watts}, hierarchical organization~\cite{ravasz2003hierarchical}, and efficient navigability~\cite{boguna2009navigability}. Therefore, understanding and analyzing them would require an expansion of standard graph models towards more specific and advanced models and methods.

We can think of the brain as a set of coarse-grained anatomical regions, connected by neuronal pathways, whose density and direction can be mapped comprehensively and non-invasively with neuroimaging techniques, namely magnetic resonance imaging (MRI). The resulting graph encodes this information in the edge weights and represents the brain's structure. Mapping brain functions to a graph, instead, involves the recording of brain activities, in the form of blood oxygen level (fMRI) or time-varying electrical or magnetic cortical signals (EEG, MEG). Here, nodes represent the recording sites and the edges encode some measure of correlation between signals. The resulting structural and functional brain networks are the objects of the so-called network neuroscience~\cite{bassett2017network}, whose main challenges are understanding the emergence of complex behavior and cognition and detecting and characterizing brain diseases~\cite{stam2014modern}.

Indeed, the presence of a neurological disease is usually reflected by alterations of local and global network properties. This has been shown for a wide variety of diseases and disorders like Alzheimer's disease~\cite{zhao2012disrupted,tijms2013alzheimer,dai2019disrupted}, stroke~\cite{adhikari2017decreased}, schizophrenia~\cite{bassett2008hierarchical}, and epilepsy~\cite{article_bdd, intra_eeg_seizure_outcome}. Comparing brain networks is a crucial task in many network neuroscience applications~\cite{mheich2020brain}. Some diseases can be seen as a network anomaly or perturbation, and the comparison of brain data from a single patient over time (or between groups of diseased and healthy subjects) is the way one can detect and localize the onset of the disease, and the starting point to predict future developments~\cite{lehnertz2023epileptic}. In recent years, for instance, brain network analysis has been applied to improve presurgical planning and to predict the postsurgical outcome of patients~\cite{intra_eeg_seizure_outcome}.

Network geometry, particularly hyperbolic geometry, has recently gained attention as a plausible latent geometric model of complex networks, from which typical properties such as a heterogeneous degree distribution and strong clustering can naturally be induced by the metric structure~\cite{krioukov2010hyperbolic}. At the same time, the approach of projecting the network onto a geometric space, called network embedding, has been used to produce highly informative low-dimensional representations of networks. 
Embedding networks in Euclidean spaces is a current approach to study and compare their connectivity structure~\cite{von2007tutorial}. However, such embeddings often require the use of high dimensions and fail to encompass some graph topological properties, such as its hierarchical structure. In contrast, non-Euclidean mapping methods often require a lower number of dimensions to accurately embed graphs and better reflect their organization~\cite{whi2022hyperbolic}. 

The hypothesis is that complex network analyses based on these representations can convey additional information on the complex system at a much lower computational cost~\cite{xu2021understanding}. When applied to structural brain networks from healthy subjects, embedding in hyperbolic space revealed a multiscale architecture of connectivity and an accurate clustering of anatomical regions in the hyperbolic disk~\cite{zheng2020geometric}. However, in the case of studying neurological diseases, this method is still rarely used. A recent study on functional brain networks (obtained from magnetoencephalography data) shows its potential in diagnosing early Alzheimer's states compared to network analysis using connectivity measures alone~\cite{baker2023hyperbolic}. However, we are not aware of any studies applying the hyperbolic embedding to epilepsy.

In this work, we study the advantage of non-Euclidean (hyperbolic) representation of networks to identify and characterize local connectivity perturbations. For this, we introduce two scores for comparing a perturbed network with its original version, the details of which are set out in sections \S \ref{subsection:method1} and \S \ref{subsection:method2}. We assess them on the localization of single-node perturbations on synthetic networks (section \S \ref{subsec:result-synthetic}). We finally apply them to localize disrupted areas following an epilepsy surgery (section \S \ref{sec:results-brain_net}).

\section{Methods}
\subsection{Network embedding in hyperbolic spaces}
\label{subsection:hypembedding}
Hyperbolic embedding is a non-linear method that takes into account the hierarchical structure of the graph (if any), and depicts both the hierarchies and the similarity of the nodes in the embedding~\cite{poincare_fb}. Embedding in hyperbolic space replicates properties of the graph, such as sparseness, self-similarity, scale-free degree distribution, small-world characteristics, and community structure, that embedding in Euclidean spaces cannot capture~\cite{linkPredHyp2020}. Hyperbolic embeddings have been proven to be useful for community detection and link prediction, as the embedding captures the similarity between network nodes in a graph~\cite{coalescent2017, linkPredHyp2020}. Therefore, we chose hyperbolic space, instead of Euclidean one, for the network embedding. This decision allowed us to capture the properties of the original network in a lower-dimensional space and quantitatively determine local perturbations to the connectivity structure.

Hyperbolic geometry studies spaces with a constant negative curvature $K$ that do not conform to Euclidean geometry. When using hyperbolic embedding, the graph is projected onto a hyperboloid that can be further projected onto a 2-dimensional hyperbolic space model like the Poincaré disk or the so-called Klein model disk. 
Current methods to embed networks in a hyperbolic disk essentially belong to two families~\cite{Saxena2020hypembsurvey, poincare_fb}: generative model-based (e.g. Mercator~\cite{GarciaMercator1019}) and machine learning-based. Here, we project our networks directly onto the 2D Poincaré disk using the coalescent embedding method~\cite{coalescent2017}, which belongs to the second family, preferring it for its remarkable versatility and computational speed.

\begin{figure*}[!htbp]
    \centering
  \includegraphics[width=17cm]{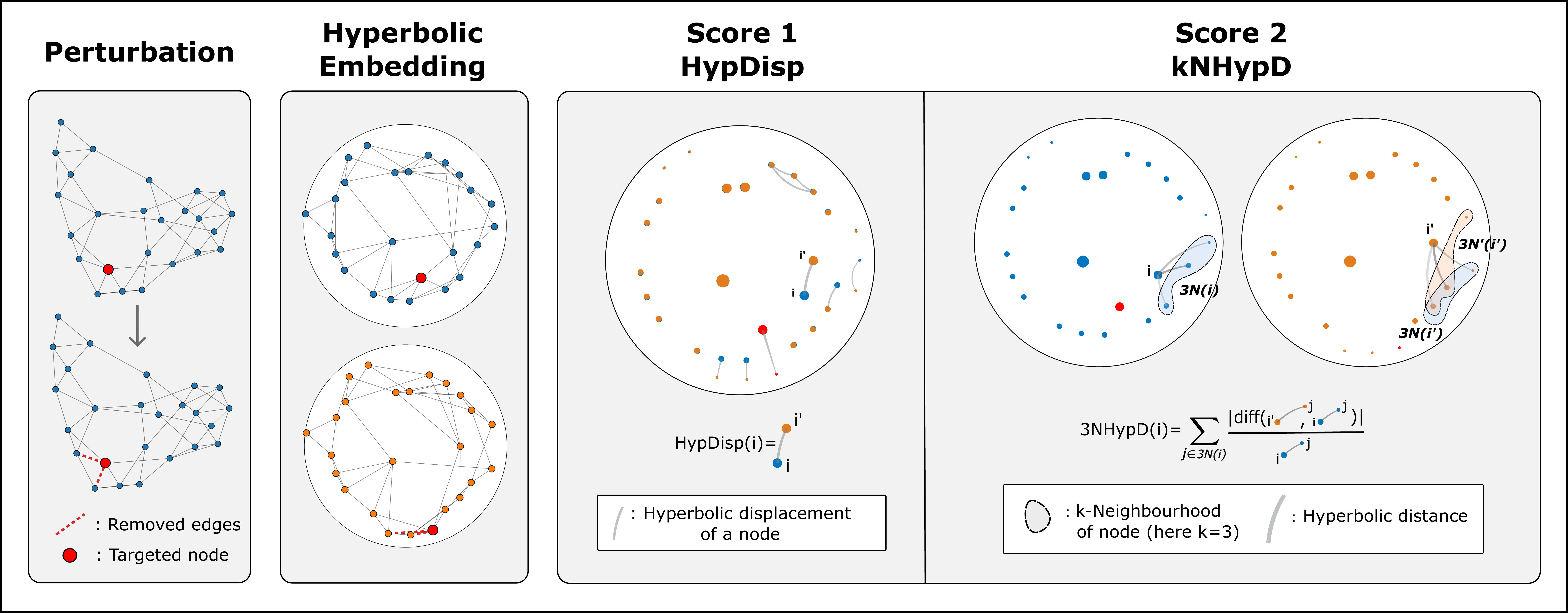}
  \caption{Pipeline for the calculation of hyperbolic perturbation scores. From left to right: network perturbation, mapping of nodes onto the Poincaré disk (lines in gray indicate the links in the original graph), alignment and comparison of original and perturbed network embedding with the HypDisp (score 1) and kNHypD (score 2) scores. In the right plots, gray curved lines represent the geodesics in the hyperbolic disks.}
  \label{fig:hyperbolic-perturbation-detection-method}
\end{figure*}

At a glance, the algorithm assigns radial coordinates to the nodes, where the radius denotes the centrality of the node, and the angle between them represents their topological proximity. In a preprocessing step, the weight of edges in the connectivity graph can be adjusted using a repulsion-attraction rule that prioritizes edges with a higher role in information transmission~\cite{coalescent2017}. The resulting matrix is then projected onto a 2D space with a nonlinear dimension reduction method, such as Isomap or Laplacian eigenmaps~\cite{von2007tutorial}. The nodes are adjusted equally in the polar coordinates by modifying the angular coordinates while maintaining the angular order of the nodes. The final step involves modifying each radius based on its rank, which is given by their degree. Within the disk, the angular distance between two nodes determines their level of similarity, according to a certain definition: for example, the number of neighbors they have in common. Additionally, each node's degree centrality is proportional to its distance from the disk's center: the closer to the center, the more central to the network. An illustration of  the hyperbolic embedding of a graph is given in the left frames of Figure \ref{fig:hyperbolic-perturbation-detection-method}. 
For the results presented in this paper, we used the dimensional reduction technique of Isomap, as it is globally optimized and preserves the global network structure by considering the shortest path distances between nodes~\cite{von2007tutorial}.

Proximity relations in the latent embedding space thus encode the importance of the interaction between nodes. In the case of the brain networks, where nodes correspond to anatomical or functional regions of interest, similar nodes that share, for instance, a high number of topological neighbors, likely belong to the same anatomical cluster and have similar characteristics as computation units in the brain. While nodes with higher centrality usually have a role of integration in the system, interacting with a larger number of anatomical regions. In the hyperbolic embedding, this is encoded in the condition that higher degree nodes are embedded to more central positions in the disk, where the hyperbolic metric naturally ensures that they are typically close, with respect to a more peripheral node, to a larger number of nodes (given that hyperbolic geodesics between peripheral points deviate toward the center).

\subsection{Perturbation detection from network embeddings}
The aim is to characterize the structural changes between the original and the perturbed network from the corresponding displacement of nodes in the hyperbolic disk. 
We quantify these changes locally by means of two scores, based on the computation of only a limited number of hyperbolic distances relative to a given node and its local neighbors. This means that once the embedding coordinates are known, the computation of the scores is relatively fast. The hyperbolic distance $\mathrm{dist}_{hyp}(i,j)$ between each pair of nodes $i$ and $j$, assigned with radii $(r_i, r_j)$ and angles $(\theta_i, \theta_j)$ at coordinates $(r_i,\theta_i)$ and $(r_j,\theta_j)$ in the Poincaré disk, is computed according to the hyperbolic law of cosines \cite{Kitsak2020hypdist}:

\begin{equation}   
\label{eq_hyp_distance}
    \cosh \mathrm{dist}_{hyp}(i,j) = \cosh r_i \times \cosh r_j - \sinh r_i \times \sinh r_j \times \cos(\pi-\vert \pi-\vert \theta_i-\theta_j\vert \vert )
\end{equation}

The complete pipeline for scores calculation is shown in Figure~\ref{fig:hyperbolic-perturbation-detection-method}. 

\subsubsection*{Score 1: Hyperbolic displacement }
\label{subsection:method1}
We define HypDisp as the displacement of a node in the hyperbolic disk caused by a perturbation. To estimate this, we embed both the initial and perturbed networks in hyperbolic space using the technique described above in Section \S~\ref{subsection:hypembedding}. Next, we align the two embeddings, as very often, a small  perturbation may introduce a random angular offset to the nodes' positions in the hyperbolic disk. A weak perturbation may thus result in two embeddings with the same similarities between nodes but with a different structure 
regarding the coordinates~\cite{gursoy2023alignment}. The alignment is achieved by rotating the perturbed embedding until the minimum sum of HypDisp, calculated between all the nodes of the whole perturbed and original embedded networks, is reached. After aligning both embeddings, a hyperbolic displacement is calculated for each node (see Figure~\ref{fig:hyperbolic-perturbation-detection-method}).  For each node $i$, its positions resulting from the original and perturbed networks are $Pos_o(i)$ and $Pos_p(i)$, respectively. The score $\mathrm{HypDisp}(i)$ attributed to the node $i$ is then :

\begin{equation}
    \mathrm{HypDisp}(i) = \mathrm{dist}_{hyp}\left(Pos_o(i), Pos_p(i)\right)
\end{equation}

where $\mathrm{dist}_{hyp}\left(X_1, X_2\right)$ is the hyperbolic distance between the positions $X_1$ and $X_2$ on the Poincaré disk (see Eq.~\ref{eq_hyp_distance}). 

\subsubsection*{Score 2: k-Neighborhood Hyperbolic Distortion}
\label{subsection:method2}

We define the $k$-Neighborhood of node $i$ relative to the hyperbolic embedding of network $G$ as the group of $k$ nodes at a smaller hyperbolic distance  from $i$, and we call it $kN(i)$. It represents the nodes with which node $i$ has a stronger interaction, as explained in Section \S \ref{subsection:hypembedding}.

To compute the hyperbolic distortion, we consider the same group of nodes in the embedding relative to the perturbed network $G'$, denoted as $kN(i')$, and quantify the change in 
 their hyperbolic distances from $i'$ due to the perturbation. We call $i'$ the node corresponding to $i$ in the perturbed network $G'$. We note that $kN(i)$ and $kN(i')$ indicate the same set of nodes, but in two different embedding configurations, meaning they will generally have different coordinates. We also specify that the set of nodes nearest to $i'$ in the embedding space of $G'$, $kN'(i')$, is generally different from $kN(i')$. These concepts are illustrated on the right side of Figure \ref{fig:hyperbolic-perturbation-detection-method}.

Let's explain how the score works in the specific: for each node $i$ we compute the set of distances $\{\mathrm{dist}_{hyp}(i,j), j \in kN(i)\}$ in the embedding space of the reference network $G$. We compare these values to $\{\mathrm{dist}_{hyp}(i',j), j \in kN(i')\}$ by estimating the score function:
\begin{equation}
	\mathrm{kNHypD}(i) = \mathlarger{\mathlarger{\sum}}_{j \in kN(i)} \frac{\left|\mathrm{dist}_{hyp}(i',j)-\mathrm{dist}_{hyp}(i,j)\right|}{\mathrm{dist}_{hyp}(i,j)}
\end{equation}

\begin{figure*}
    \centering
  \includegraphics[width=15cm]{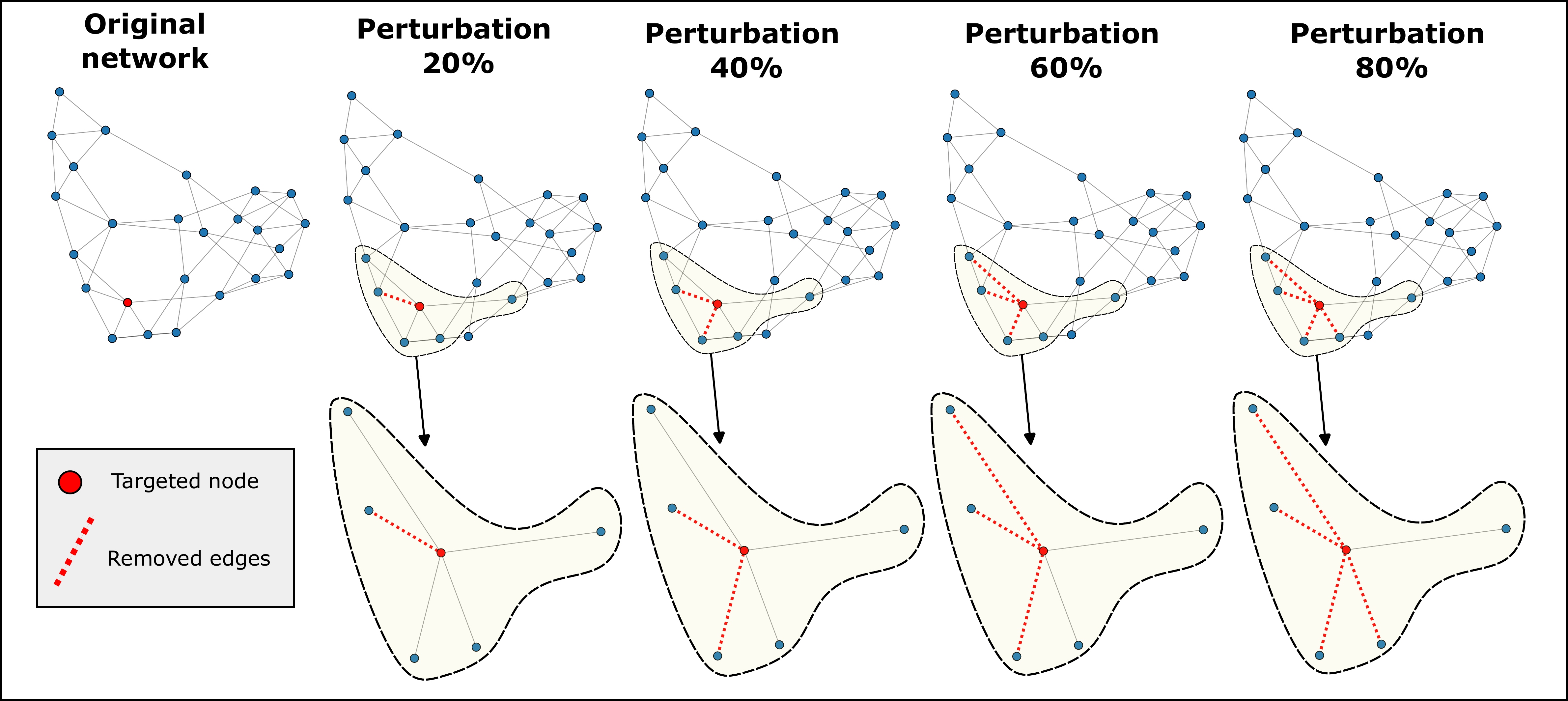}
  \caption{The network perturbation method. The method involves randomly selecting a target node and then removing some of its edges based on the desired level of perturbation.}
  \label{fig:perturbation-method}
\end{figure*}

The $\mathrm{kNHypD}(i)$ score reflects the degree of distortion in the $k$-Neighborhood of node $i$ due to a local perturbation of its connectivity. By adding the absolute value, we model a positive contribution to the score in the case where a distance within the $k$-Neighborhood is reduced by the perturbation. This may occur due to a rewiring process: increasing the degree of a node can push it toward the center of the disk, or increasing the connectivity between two nodes can bring them closer. Dividing the difference by $\mathrm{dist}_{hyp}(i,j)$ gives a higher weight to nodes with which $i$ has a stronger interaction. Normalizing by a quadratic factor $\mathrm{dist}^2_{hyp}(i,j)$ can ensure convergence of the score for $k \rightarrow N$ and $N \rightarrow \infty$.
A strong point of this measure is that it is invariant under rotations and reflections of the embedding space; therefore, no prior alignment of the two embeddings to be compared is necessary. 

The best choice of $k$ in the estimation of kNHypD depends upon the data. As for other non-parametric estimators in statistics~\cite{berrett2019efficient}, the variance of the estimator appears to be high for small values of $k$. Although increasing the size of the neighborhood reduces the variance, it may also yield very coarse estimates of kNHypD thus increasing the bias. Moreover, the stability gained by using higher values of $k$ can be overshadowed by the complexities involved in computing a large number of distances. Overall, based on the score's performances across the different networks considered, it is recommended to choose a neighborhood size between 10\% and 50\% of the network size.

\section{Data}
\subsection{Synthetic data}
We first tested our approach on simulated networks using two models: the Watts-Strogatz model~\cite{Watts}, otherwise referred to as the small-world networks, and the Barab\'asi-Albert model~\cite{barabasi}, which consists of networks with a power-law degree distribution. The Watts-Strogatz model yields a more homogeneous connectivity, while the Barab\'asi-Albert model produces a more heterogeneous structure and includes hubs. 

\subsubsection*{Simulation of network perturbations} \label{subsection:perturbation-method}
We examined the change in the embedding configuration of networks under a ``disconnection'' perturbation, where nodes are retained (not removed, relabeled, or isolated), but some of their links are cut. This type of disruption mirrors what brain networks experience during epilepsy surgery before any connectivity rewiring mechanism begins. Specifically, we designed a random perturbation method  that allows for control over the number of target nodes and the size of the perturbation. This method randomly cuts a percentage of links among those available for each target node(see Figure~\ref{fig:perturbation-method}). To ensure method compatibility with real brain networks (see below), we selected synthetic networks with similar size and density: $N=150$ nodes, parameters $k=20$ and $p=0.1$ for Watts-Strogatz, and $m=8$ for the scale-free networks. In our simulations, we excluded realizations with disconnected nodes in order to maintain a connected network.

\begin{figure*}
    \centering
  \includegraphics[width=17cm]{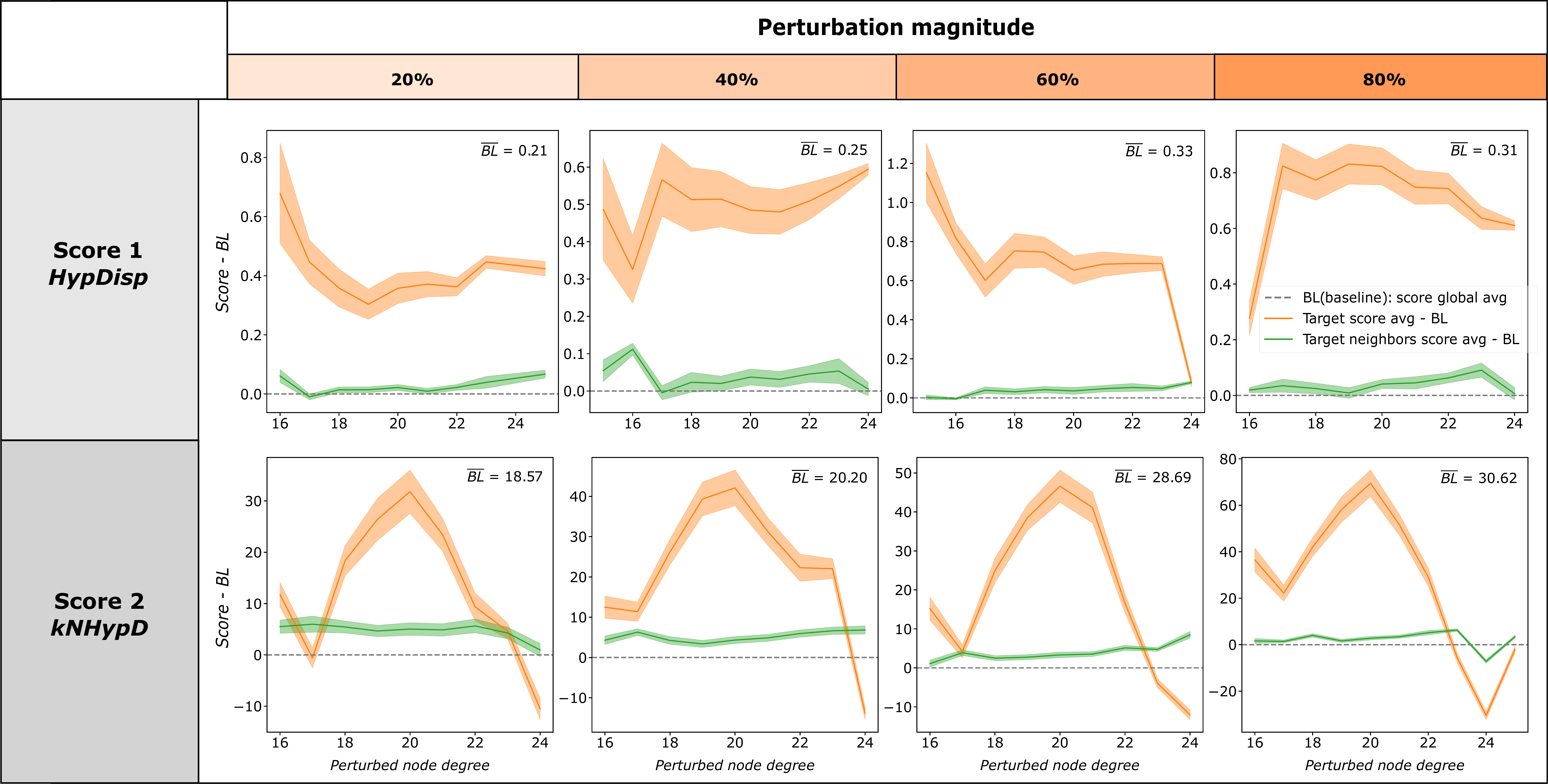}
  \caption{Results on synthetic networks. HypDisp and kNHypD scores are computed for each node taken as target in a Watts-Strogatz network with $N=150$ nodes, a lattice parameter of $20$ and $0.1$ rewiring probability. We visualize the average values per degree class, measured at the target node itself (in orange), and at the level of its nearest topological neighbors (in green), both relative to the global average over the network, the baseline (dashed grey line). In each subplot $\overline{BL}$ indicates the total average of the scores over all the network's nodes. The bandwidths indicate the $95\%$ confidence interval of the mean. For the estimation of kNHypD, we set the number of neighbors equal to 10.}
  \label{fig:WS-results}
\end{figure*} 

\subsection{Clinical data}
\label{clincal_data}
Finally, we assessed our scores using anatomical brain connectivity networks derived from patients who underwent epilepsy surgery. We examined the brain networks of 51 patients, estimated using T1-weighted structural and diffusion-weighted MRI scans before and after surgery. The pre- and post-operative networks were obtained from an open dataset associated with the research paper by Sinha et al.~\cite{article_bdd}. The dataset included 51 individuals who underwent epilepsy surgery, from which 30 patients underwent surgery for a left temporal lobe epilepsy (TLE), and 21 for a right temporal lobe epilepsy. Within each category, participants were further divided based on post-surgical results: those with positive surgical outcomes (seizure-free after surgery) and those with negative outcomes (not seizure-free after surgery). The surgical procedure involved removing brain tissue located in the presumed epileptic zone (specifically, in the temporal lobe) to prevent seizures. In graph theory, this corresponds to cutting a subset of edges within a localized section of the network. The brain networks before the surgery have the following properties: a mean degree of $28 \pm 2$ nodes, a standard deviation of the degree distribution of $12 \pm 1$, a mean clustering coefficient of $0.57 \pm 0.02$. After surgery, the perturbed networks display a mean degree of $25\pm 2$, a standard deviation of the degree distribution of $11 \pm 1$, and a clustering coefficient of $0.55 \pm 0.03$. Brain networks have $N=114$ nodes and there are no disconnected nodes after surgery.

\section{Results}
\subsection{Synthetic networks}
\label{subsec:result-synthetic}
When the network's connectivity is locally disrupted, we anticipate that this will be apparent in a change in the proximity relationships of the disrupted node within its embedding space, and we expect our scores to detect this change.   We hypothesize 
that the geometric distortion will increase statistically with the size of the disruption, both locally and globally.

To test this hypothesis, we apply the perturbation method described in \S \ref{subsection:perturbation-method} to selected synthetic connectivity models —specifically, small-world (Watts-Strogatz) and scale-free (Barab\'asi-Albert) networks. To avoid specific perturbations from influencing the results, we repeatedly apply random perturbations to a single node a number of times that is proportional to the number of possible distinct perturbations. We estimate this number as $N_p = \binom{\bar{k}}{\round{\bar{k} \times m}} \cdot N$, where $N$ is the network's size, and the other factor is an estimate of the possible perturbations for a node with a degree equal to the average degree $\bar{k}$. In the binomial coefficient estimation, $m$ indicates the perturbation size as a fraction of links to cut, and $\round{x}$ denotes the nearest integer to $x$.

\begin{figure*}[t]
    \centering
  \includegraphics[width=17cm]{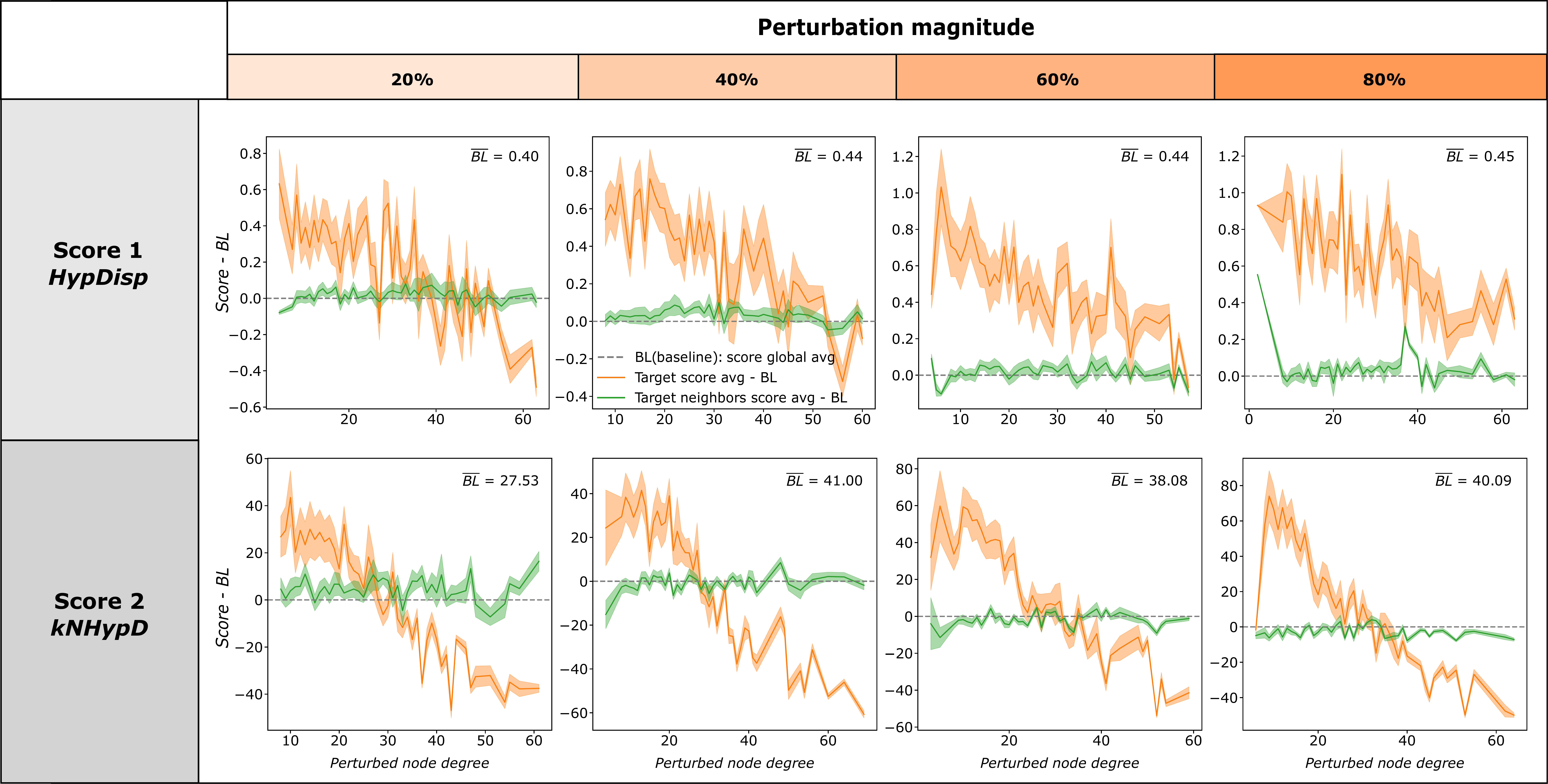}
  \caption{HypDisp and kNHypD scores are computed for each node taken as target in a Barab\'asi-Albert model with $N=150$ nodes and $m=8$. Same stipulations as in the caption of Figure~\ref{fig:WS-results}}
  \label{fig:BA-results}
\end{figure*}

We uniformly sample within the degree domain, allowing nodes with high degree $k \gg \bar{k}$ —and thus a higher number of possible perturbations— to be sampled more frequently. Additionally, we choose the sample size to be a fraction $\nicefrac{N_p}{\alpha}$ with $\alpha > 1$, ensuring both the statistical significance of the analysis and the smoothness of the computation. For a given network model (small-world or scale-free), we also average over $n_G$ random network realizations.

In the trivial case where a perturbation of zero magnitude is applied, the embedding configuration of $G$ and $G'$ are identical. As expected, both the kNHypD and HypDisp scores yield a value of~$0$ uniformly across the network. Instead, the embedding configuration always changes when a non-null perturbation is applied.

To quantify how much the embedding configuration is distorted locally and globally, for each perturbation we compute: i) the two geometric scores at the target node; ii) the average value of the scores relative to the nearest neighbors of the target node, i.e. the nodes sharing an edge with the target node (they are just after the target node, the most physically affected by the perturbation). 
iii) the baseline (BL) defined as the average of the perturbation scores across the entire network. We then divide the results obtained for the $n_p$ perturbations over $n_G$ random network realizations based on the degree of the perturbed node. This enables us to observe the perturbation as a function of degree. 
Finally, for each degree class, we compute the average values of (i) with respect to the baseline (iii), indicated in the figure as ``Target score avg - BL''. Similarly, we obtain the average value of (ii) with respect to BL, referred as ``Target neighbors score avg - BL''. The measure $\overline{BL}$ reported in each frame indicates the value of the baseline further averaged over all the degree classes. 

The results for the small-world network model are shown in Figure~\ref{fig:WS-results}. The HypDisp scores at the target nodes are significantly higher than the baseline scores, practically for all nodes' degree classes. There is only a slight convergence of the target node's score and the score from its second-order neighbors at the extremes of the degree domain, but there is never a reversal of the trend or an overshoot of the baseline score. The kNHypD score also correctly identifies a perturbation, except in the extremes of the nodes' degree domain. If the targeted node has few connections, the perturbation cuts too few links relative to the original link density, and the score may not be sensitive enough. On the other hand, if the targeted node has more links, the perturbation results in a delocalization of the perturbation effect over many areas of the network, which can result in a target score comparable to the baseline.

The values of the average baseline $\overline{BL}$ —defined here as the global average of the scores over all the network's nodes and over all network realizations and perturbations— suggest that the magnitude of the perturbation definitely plays a role in the performance of the scores. As anticipated in the beginning of this section, a stronger local perturbation induces a higher distortion over larger parts of the network, which is reflected in increasing values of $\overline{BL}$ (this effect is reported in Section 2 of the Supplementary Material). From the bottom plots in Figure~\ref{fig:WS-results}, we also observe that a perturbation of higher magnitude yields a larger difference between the target kNHypD score and baseline.

The results for the scale-free network model are presented in Figure~\ref{fig:BA-results}. For HypDisp score, results indicate that the localization power becomes more stable as the perturbation magnitude increases (top plots in Figure \ref{fig:BA-results}). Regarding the kNHypD score, there is still a notable increase in the peak distance from the baseline, but it fails to identify the perturbation when high-degree nodes or hubs are targeted. Since this model has a power-law degree distribution, perturbing a hub leads to disturbing a larger number of distant regions of the network, decreasing the difference between the error of the perturbed node and the baseline. As shown in Figure~\ref{fig:BA-results}, this effect appears for both scores and is more pronounced for weak perturbations.

\begin{figure*}[t]
    \centering
  \includegraphics[width=17cm]{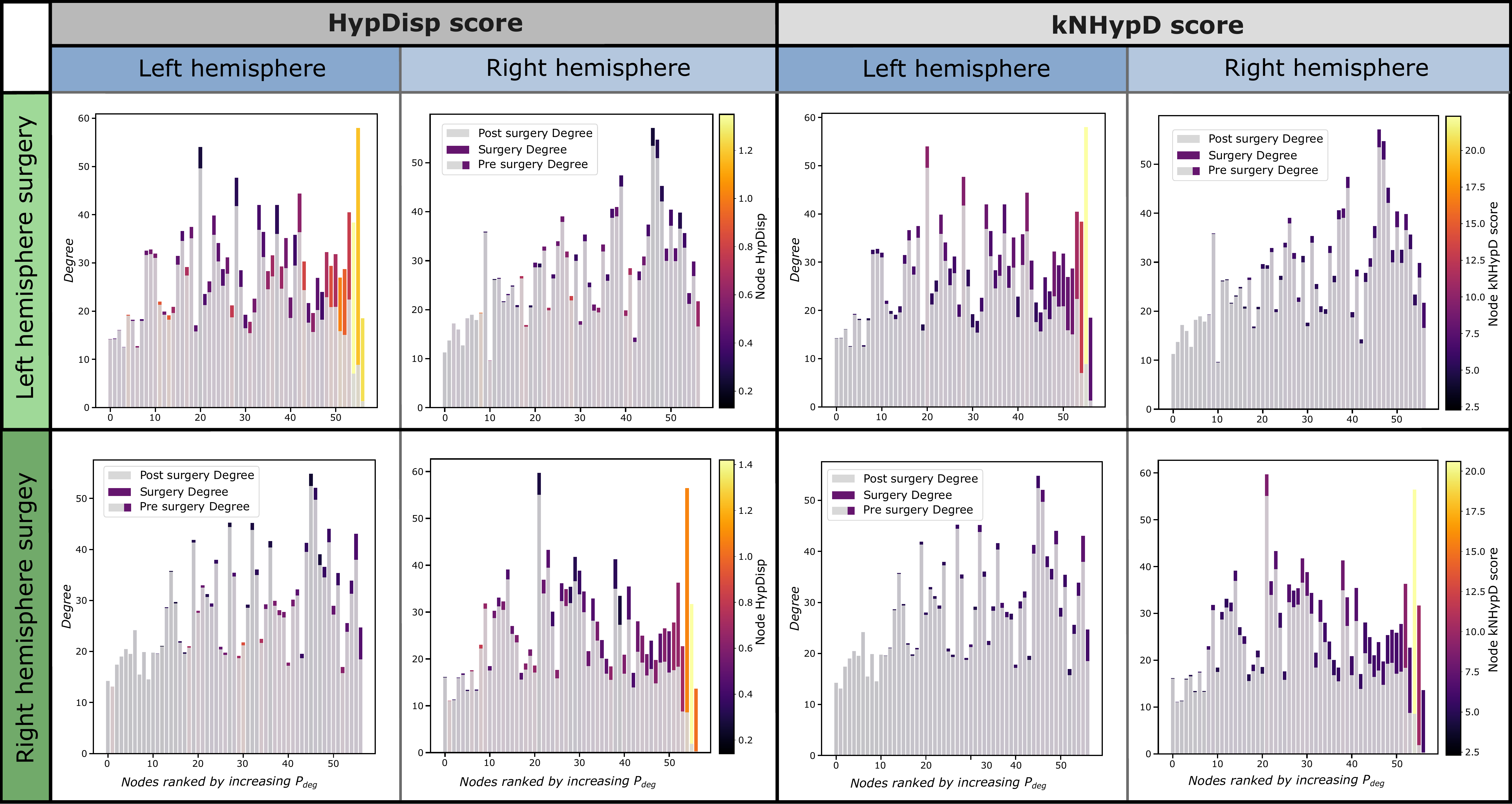}
  \caption{Subplots in the first row show results from patients operated in the left hemisphere, whereas those operated in the right hemisphere are displayed in the second row. The first two columns display results from HypDisp score, whereas results from kNHypD score are shown in the 3rd and 4th columns. For each score, nodes were separated into the left hemisphere (left subplot) and  right hemisphere (right subplot). In the different subplots, the nodes were assigned a ranking based on increasing perturbation degree $P_{deg}$ from left to right. Color bars encode the mean score of the node across patients. The total height denotes the node's mean pre-surgery degree, with mean post-surgery degree is indicated by a shaded gray section (always lower than the pre-surgical values).}
  \label{fig:result-brain-network-alumettes}
\end{figure*}

For both network models, the scores calculated from the second-order neighbors of targeted nodes are similar to those obtained from the baseline, suggesting a good localization power, as these nodes are practically unaffected by perturbations. Strong local correlations of nodes (assortativity), however, could have an effect and induce a bias in the scores of these ``second-order target nodes''. Interestingly, at the average degree kNHypD scores show a peak in Watts-Strogatz networks and a crossing with the baseline in the Barabási-Albert networks. In contrast, no clear trend was found for the kNHypD score as a function of the target node's clustering coefficient (results in the Supplementary Material). Therefore, we can conclude that: i) the proposed perturbation scores have good localization and a small bias; ii) the performance improves with perturbation magnitude; and iii) the performances of the kNHypD score depend on degree heterogeneity. 

In the calculation of the kNHypD score we have taken into account the displacement of the 10 nearest geometric neighbors of each node, i.e. $k=10$. In the Supplementary Material, you can find the results for two extreme values of this parameter ($k=1$ and $k=100$).  

\subsection{Brain networks}
\label{sec:results-brain_net}

To illustrate the effectiveness of our scores on real data and compare their performances, we used them on brain network connectivity perturbations due to epilepsy surgery (refer to \S \ref{clincal_data} for further details). For  illustration purposes, here we considered patients with a positive surgical outcome. For these patients, we compared two networks: the pre- and post-surgery networks. The network after surgery is a modified version of the original network, with fewer edges specifically removed in the surgery area. A way to assess the perturbation of node $i$ due to surgery is the perturbation degree, $P_{deg}(i)$, defined as $P_{deg}(i) = \nicefrac{(D_{pre}(i)-D_{post}(i))}{D_{pre}(i)}$, where $D_{pre}(i)$ and $D_{post}(i)$ denote the pre- and post-surgery degree of the i$^{th}$node. When $P_{deg}(i) \rightarrow 0$, the node's connectivity is slightly affected by the surgery. In contrast, when $P_{deg}(i) \rightarrow 1$, the node is more likely to become disconnected and significantly impacted. In both groups (left and right hemisphere surgery), three nodes out of the 114 have a perturbation magnitude greater than $0.8$, while 109 nodes have a perturbation degree below $0.4$. The average perturbation magnitude across all patients and nodes is $0.118 \pm 0.032$.

As our scores aim to locate the disturbed region of a network, we anticipate a correlation of our scores with the perturbation degree distribution. Results shown in Figure~\ref{fig:result-brain-network-alumettes} confirm that nodes exhibiting the highest scores (encoded by colors that tend to be clearer), correspond to the nodes exhibiting the highest perturbation degree. When surgery affects brain regions with few connections, both perturbation scores yield very low values. These findings have been consistently observed in both hemispheres. A very important point is that nodes in the surgical hemisphere clearly display a stronger perturbation score than those located on the contralateral hemisphere, independently of the perturbation degree. 

\begin{figure*}[!t]
    \centering
  \includegraphics[width=17cm]{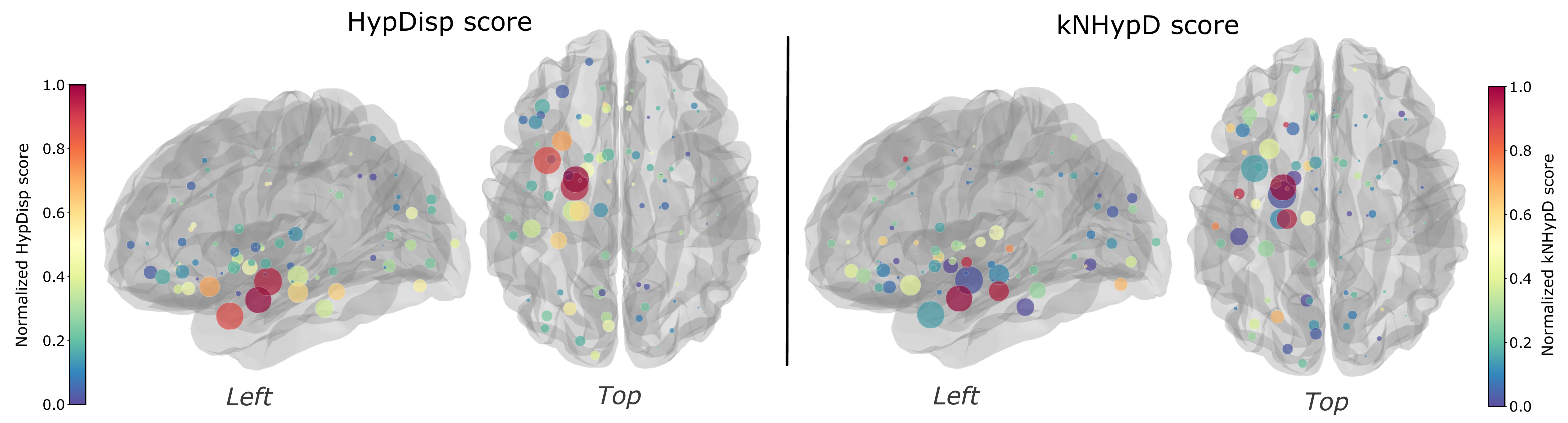}
  \caption{Left and top views of brain network perturbations in epileptic patients who have undergone left hemisphere surgery. Each node's size corresponds to the surgery perturbation rate (rate of removed connections) of the node. The color of nodes represents its score by either HypDisp (score 1) or kNHypD (score 2). For the sake of visualisation, values are normalized between $0$ and $1$. }
  \label{fig:resul-res-brain}
\end{figure*}

The spatial mapping of brain regions' perturbation is illustrated in Figure~\ref{fig:resul-res-brain}. In both cases, one can see that nodes with the highest scores are always located in the surgically impacted region (temporal lobe). We notice, however, that the perturbation score calculated with the HypDisp score covers a large region, including all perturbed nodes; whereas the kNHypD score identifies a more reduced number of brain areas within the region directly affected by the surgery.

\section{Discussion}
The characterization and localization of network perturbations have applications in various domains such as anomaly detection (in disease networks, computer networks, etc.) and brain surgery characterization. This study introduces two scores based on hyperbolic embedding to characterize and localize local network perturbations. Results from synthetic networks validate the proficiency of the proposed scores in identifying perturbed nodes. To showcase their practical utility, the scores are applied to real brain networks from epileptic patients who underwent surgery. The results reveal that both scores successfully pinpointed the surgical area as the most perturbed region. Notably, the HypDisp score identified significant perturbations in a broader brain zone beyond the surgical region, whereas the kNHypD score highlighted perturbations in a more localized area.

The novelty of our scores resides in an explicit analysis in a latent hyperbolic space. We expect that the use of these low-dimensional and highly informative representations will be crucial for brain network studies beyond standard connectivity measures and Euclidean geometry. 

For this study, the coalescent embedding method~\cite{coalescent2017} has been selected from a range of available methods~\cite{Saxena2020hypembsurvey, poincare_fb}. This method encompasses various benefits, such as \textit{i)} the absence of a minimization step that prevents errors introduced by local minima and also increases reproducibility (the same graph used as input will always produce the same embedding), and \textit{ii)} a shorter computational time with respect to methods involving a likelihood maximization like Mercator~\cite{GarciaMercator1019}. The latter, based on a generative model, has the advantage of producing more meaningful representations. This could allow, for instance, to generate synthetic networks with similar properties to the input network. However, for the purposes of this paper, we have decided to prioritize speed and versatility over this feature. A comparison of the computational time of embeddings using coalescent and Mercator methods can be found in the Supplementary Material. 

Similarly, we notice that other embedding techniques could potentially be combined with the proposed perturbation scores, and that alternative dimensional reduction methods (e.g., Laplacian Eigenmaps~\cite{von2007tutorial}) or hyperbolic mappings (e.g., Mercator~\cite{GarciaMercator1019}, HyperMap~\cite{papadopoulos2014network}, or Hydra~\cite{keller2020hydra}, among others) might be worth considering for the representation of brain networks~\cite{whi2022characteristic}.

In conclusion, the article presents two scores that efficiently characterize network perturbations of the type link removal. Future work could explore the characterization of other types of perturbations, such as rewiring. Our approach may provide new insights in the study of perturbations of connected systems including social, technological or biological networks.

\section*{Supplementary Material}
See the supplementary material for: (1) the dependency of the kNHypD score on the value of parameter $k$, (2) the evolution of the mean baseline ($\overline{BL}$) with the perturbation degree, (3) a comparison of computation times, as a function of network size, of  the Mercator and the coalescent hyperbolic embedding method, and (4) the dependency of the kNHypD score on the clustering coefficient.

\section*{Acknowledgements}
Authors are grateful to Steven Rico for useful and valuable discussions during the preparation of the study. A.L. was supported by the doctoral school EDITE from Sorbonne University. M.G. acknowledges doctoral support from the Ecole Normale Sup\'erieure de Lyon. We thank the two anonymous referees for helpful comments.

\section*{Author declarations}
\subsection*{Conflict of Interest}
The authors have no conflicts to disclose.
\subsection*{Author Contributions}
A.L. and M.G. contributed equally to this work.\\

\textbf{A. Longhena} Conceptualization (equal); Formal analysis (equal); Methodology (equal); Visualization (equal); Writing/Original Draft Preparation (equal); Writing – review \& editing (equal).
\textbf{M. Guillemaud} Conceptualization (equal); Formal analysis (equal); Methodology (equal); Visualization (equal); Writing/Original Draft Preparation (equal); Writing – review \& editing (equal).
\textbf{M. Chavez} Conceptualization (equal); Supervision (lead); Writing – review \& editing (equal).

\section*{Data Availability}
The brain network data that support the findings of this study are publicly available in the supplementary material of this \href{https://www.neurology.org/doi/10.1212/WNL.0000000000011315}{article}~\cite{article_bdd}.

\bibliography{main_X_HPL_v4}

\end{document}